\documentclass[10pt, reqno, a4paper]{amsart}
\pdfoutput=1

\usepackage{amsthm, amsmath, amssymb}
\usepackage{graphicx}
\usepackage{caption}
\usepackage{subcaption}
\usepackage{wrapfig} 
\usepackage{hyperref}
 
\newcommand{\pardertwo}[1][r]{\ensuremath{\frac{\partial^2}{\partial #1^2}}}

\newcommand{\N}{\mathbb{N}}
\newcommand{\R}{{\mathbb{R}}}
\newcommand{\Z}{{\mathbb{Z}}}
\newcommand{\dd}{{{\rm d}}}
\newcommand{\ii}{{\rm i}}
\newcommand{\cf}{\emph{cf.}}
\newcommand{\ie}{{\emph{i.e.}}}
\newcommand{\eg}{{\emph{e.g.}}}

\begin{document}
 
\title{Energy spectrum of the hydrogen atom in a space with one~compactified extra dimension, $\R^3 \times S^1$}

\author{Martin~Bure\v s}
\address[Martin~Bure\v s]{
Institute for Theoretical Physics and Astrophysics,
Masaryk University, Kotl\'a\v{r}sk\'a 2, 61137 Brno, Czech Republic
}
\email{bures@physics.muni.cz}
\thanks{The author would like to thank Rikard von Unge who suggested this problem and who provided insight and expertise that greatly assisted the research. 
	Gratefully acknowledged is the help and guidance of Jean-Marc Richard and Maurice Kibler who gave the author the possibility of conducting a similar research at IPNL Lyon. 
        JMR helped to improve with valuable comments and suggestions also this manuscript and provided his Mathematica code for the Wronskian method.
	The work was supported by the Czech government grant agency under contract no. GA\v CR 202/08/H072.
}


\date{\today}

\maketitle

\begin{abstract}
	We investigate the consequences of one extra compactified dimension for the energy spectrum of the non-relativistic hydrogen atom with a potential defined by Gauss' law, \ie~proportional to~$1/|x|^2$ in non-compactified 4d space. The calculations were performed numerically by diagonalizing the Hamiltonian in two different sets of basis vectors. The energy levels and electron probability density are plotted as a function of the compactification radius. The occurrence of several physical effects is discussed and interpreted. 
\end{abstract}


\section{Introduction}
%
In view of modern theories, aspiring to unify gravity with the remaining forces, the study of higher-dimensional models becomes very important.
Surprisingly enough, in spite of many years of intensive research, one of the basic and most interesting questions, namely what happens to ordinary atoms when more than the usual three dimensions are considered, seems not fully answered. 
 
The problem of stability of the hydrogen atom in a space with one additional spatial dimension and a potential defined by Gauss' law was subject to a recent article~\cite{Bures-Siegl-2015}. Apart from an extra spatial dimension of an infinite extent (resulting topology $\R^4$), the authors considered an extra dimension compactified to a circle (resulting topology $\R^3\times S^1$). The authors proved that the \textit{hydrogen atom in a compactified universe is stable} for compactification radii smaller than the critical value $R_{crit}=a_0/4$, where $a_0$ is the Bohr radius. Furthermore, it was demonstrated that the system possesses an infinite number of bound states, with an energy extending at least to the ground state energy of the hydrogen atom.

The purpose of the present work is twofold. Firstly, to calculate the approximate energy spectrum as a function of the compactification radius. Secondly, to discuss various effects that appear in the spectrum and in the electron probability density when the compactification radius increases. 
It is particularly important to investigate if the calculations of energy levels reveal some indication of divergent behavior of the ground state energy at compactification radii close to the critical value $R_{crit}$~\cite{Bures-Siegl-2015}. 

To the best of our knowledge of existing literature, there is a lack of results in this direction, and this important problem is still unresolved. 
Perhaps the closest attempt was that of~\cite{Floratos2010}, where the Bergmann-Frishman transformation~\cite{Bergmann1965} was used to reformulate the hydrogen problem in terms of a four dimensional harmonic oscillator. Then, the calculation of energy corrections was performed, however, only a different, simplified form of the potential than that corresponding to an extra compactified dimension, was treated eventually.

\section{Hydrogen atom in non-compactified spaces $\R^d$}
%
Some authors define the $d$-dimensional Coulomb problem through the~$1/|x|$ potential irrespective of the number of extra dimensions, \eg~\cite{burgbacher1999,Nieto1979,jaber1998}. 
Let us emphasize that we deal with the physically more justified potential as given by Gauss' law in $d$ dimensions.
Such an approach gives the potential $V_d(|x|)\sim |x|^{2-d}$ and the corresponding Schr\"odinger equation reads
\begin{eqnarray}\label{schr.eq}
\left(-\frac{\hbar^2}{2m}\Delta- \frac{e^2_d}{|x|^{d-2}} \right) \psi= E \psi ,
\end{eqnarray}
where $e_d$ is a $d$-dimensional charge.
This is an utterly different situation. A potential obtained from Gauss' law, when treated in higher dimensional spaces with \textit{non-compactified} extra dimensions, leads to instability~\cite{buechel,Freeman1969,Gurevich1971,Morales1996,Braga:2005ai,Essin2006,Andrew1990}. This result remains valid also in the case of four spatial dimensions where, however, a special treatment is required. Due to the degree of singularity, the potential energy term $1/|x|^2$ can be merged with the centrifugal term arising from radial reduction of the Schr\"odinger equation in spherical coordinates. 
This results in the existence of a critical value for the (dimensionless) coupling parameter $Z:=2me_4^2/\hbar^2$.
For $Z \in [0,Z_{crit})$, the system is stable without any bound state solutions, for $Z>Z_{crit}$, the energy of the system is unbounded from below~\cite{Bures-Siegl-2015,Krall1982,Gurevich1971}. Thus, there is no stable hydrogen atom if the extra dimension is infinite.
However, if we compactify the extra dimension below a certain critical value of the compactification radius,
then Gauss' law gives a potential that results in a stable system, as was shown in detail in~\cite{Bures-Siegl-2015}.

\section{Hydrogen atom in space with one extra compactified dimension, $\R^3\times S^1$}
%
Following~\cite{Bures-Siegl-2015}, we consider the underlying space to be $\R^3\times S^1$, where
$S^1$ is the fourth dimension $x_4$ \textit{compactified in a circle} of radius $R$.
The parametrization of $S^1$ by using an angular coordinate $\theta$ via $x_4=R\theta$ then leads to the compactified configuration space $\R^3 \times (-\pi,\pi)$.
The potential of a point charge in four dimensions, \ie~$-e_4^2/|x|^2$, where $e_4$ is the four-dimensional charge (with unit $\text{energy}^{1/2} \times \text{length}$), determines the potential in the compactified case:
\begin{eqnarray} \label{pot-1ED-sum}
	V(r,\theta)=-\sum_{n=-\infty}^{\infty} \frac{e_4^2}{r^2+R^2(\theta-2\pi n)^2},
\end{eqnarray}
where $r^2=x_1^2+x_2^2+x_3^2$ is the radius vector in three dimensions. 
This expression can be summed and written in a closed form as
\begin{eqnarray} \label{pot-1ED-closed}
	V(r,\theta)=-\frac{e^2}{r}\,\frac{\sinh{(r/R)}}{\cosh{(r/R)}-\cos{\theta}}
                  =-\frac{e^2}{r}
		   -\frac{2e^2}{r}\,\frac{e^{r/R}\cos{\theta}-1}{e^{2r/R}-2e^{r/R}\cos\theta+1},
\end{eqnarray}
where we choose the relation between three- and four-dimensional charges to be:
\begin{equation}\label{relation}
e_4^2/2R=e_3^2\equiv e^2=\hbar^2/ma,
\end{equation}
with $a$ being the Bohr radius. This relation comes from the limiting behavior of the potential at large distances. For details, we again refer to~\cite{Bures-Siegl-2015}.
Here, we will need the Fourier expansion of the potential~\eqref{pot-1ED-closed}:
\begin{equation}\label{pot.fourier}
	V(r,\theta)
	=\sum_{n=-\infty}^{\infty}  v_n(r)\, e^{in\theta},\quad
	v_n(r)=-\frac{e^2}{r} e^{-|n|r/R}, \quad \text{or} \qquad
	V(r,\theta)
	=-\frac{e^2}{r}-\frac{2e^2}{r}\sum_{n=1}^{\infty} e^{-nr/R}\, \cos{n\theta}.
\end{equation} 
Thus, the Hamiltonian of the non-relativistic hydrogen atom in a space with one compactified extra dimension $\R^3\times S^1$ is
\begin{eqnarray}\label{ham}
\hat{H}=\hat{H}_{3\text{d}} -\frac{\hbar^2}{2mR^2}\pardertwo[\theta]- \frac{e^2}{r}\sum_{k\neq 0} e^{-|k|r/R} e^{\ii k \theta},
\end{eqnarray}
where $\hat{H}_{3\text{d}}$ is the usual hydrogen atom Hamiltonian,
\begin{equation}
	\hat{H}_{3\text{d}}= -\frac{\hbar^2}{2m}\Delta_{3\text{d}}- \frac{e^2}{r}.
\end{equation}
For the following, we shall introduce the units of $e^2/2a$ for energy and $a$ for length (so that $\hbar^2/2m=1$ and the elementary charge is fixed to $e^2=2$). 
%
\section{A toy model in~$\R^3$: Coulomb potential perturbed by a Yukawa potential}\label{sec.toy}
%
Let us first illustrate the computational method on a related example. Because the potential in~\ref{ham} is given by a sum of Yukawa-type terms in the $r$-variable, we will illustrate the use of the method of Hamiltonian diagonalization on the simple case of a Yukawa potential. We will then compare the results to those obtained by a direct numerical integration of the radial equation.
We consider the Hamiltonian 
\begin{eqnarray}\label{hamtoy}
	\hat{H}=\hat{H}_{3\text{d}} -g \frac2{r} e^{-\mu r},
\end{eqnarray}
with
$ 	\hat{H}_{3\text{d}}= -\Delta_{3\text{d}}- 2/r$. Here $g>0$ and $\mu>0$ is the strenght and the range parameter, respectively.
%
\subsection{Hydrogen atom basis}\label{sec.toy.H}
%
As a first basis set, we choose the bound states of the hydrogen:%
\footnote{We use the following definition of the associated Laguerre polynomials (compatible with Mathematica): $L_n^{\alpha}=\sum_{i=0}^n\binom{n+\alpha}{n-i}\frac{-x^i}{i!}$. Then $R_{nl}$ are orthonormal.} 
\begin{equation}\label{wfctstoy}
	\langle \vec{x}|nlm \rangle = R_{nl}(r)Y_{lm}(\theta,\phi), \quad l\in\N,\, m\in\{-l,\dots,l\},\, n\in\{l+1,l+2,\dots\},
\end{equation}
where
\begin{equation}\label{rad-fct}
	R_{nl}=\frac{2}{n^2}\sqrt{\frac{(n-l-1)!}{(n+l)!}} \left(\frac{2r}{n}\right)^l e^{-r/n} L_{n-l-1}^{2l+1}\left(\frac{2r}{n}\right)
\end{equation}
are the radial function of the hydrogen atom and $Y_{lm}(\Omega)$ are the spherical harmonics.
After some calculations, we get for the matrix elements of~\eqref{hamtoy} the following expression:
\begin{equation}\label{matrixtoy}
	\langle n'l'm' |\hat{H}|nlm \rangle = \delta_{ll'} \delta_{mm'}
	\left \{ -\frac1{n^2}
\delta_{nn'} 	- M_{n,n';l}\left(g,\mu\right)  \right \}\, ,
\end{equation}
where the matrix elements $M_{n,n';l}$ are given by 
\begin{multline}\label{elementintegraltoy}
	M_{n,n';l}\left(g,\mu\right) 
	 := 2g\int_0^{\infty}\dd r r^2 R_{nl}R_{n'l} \frac{e^{-\mu r/a}}{r/a}\\
	 = \frac{g}{2}\left(\frac{4}{nn'}\right)^{l+2}\sqrt{\frac{(n-l-1)!(n'-l-1)!}{(n+l)!(n'+l)!}} 
	 \int_0^{\infty}\dd x \,x^{2l+1}   e^{-\sigma x}
	L_{n-l-1}^{2l+1}\left(\frac{2x}{n}\right) L_{n'-l-1}^{2l+1}\left(\frac{2x}{n'}\right),
\end{multline}
with $\sigma(\mu)=1/n+1/n'+\mu$. 
To evaluate the integral, we use formula~\cite[Eqn.~(14)]{Srivastava2003}:%
\footnote{\label{foot-formula} The referred formula is slightly more general. For our purpose, we can simplify it using ${}_1F_2\left(a,b;b;z\right)=(1-z)^{-a}$ to get
$\int_0^{\infty}\dd x \,x^{\alpha}   e^{-\sigma x}
L_m^{(\alpha)}(\lambda x)
L_n^{(\alpha)}(\mu x)
=
\frac{\Gamma(\alpha+1)}{\sigma^{\alpha+1}}
	\sum_{k=0}^{min(m,n)} 
	\binom{m+\alpha}{m-k}
	\binom{n+\alpha}{n-k}
	\binom{k+\alpha}{k}
        \left(\frac{\lambda\mu}{\sigma^2}\right)^k	
	\left(1-\frac{\lambda}{\sigma}\right)^{m-k}
	\left(1-\frac{\mu}{\sigma}\right)^{n-k}
$.}
 
\begin{multline}\label{elementtoy}
	M_{n,n';l}(g,\mu)
	=\frac{g}{2}\left(\frac{4}{nn'}\right)^{l+2}\sqrt{\frac{(n-l-1)!(n'-l-1)!}{(n+l)!(n'+l)!}}
        \frac{(2l+1)!}{\sigma^{2l+2}}
	\sum_{k=0}^{min(n-l-1,n'-l-1)}\binom{n+l}{n-l-1-k}\\
	\times
	\binom{n'+l}{n'-l-1-k}
	\binom{k+2l+1}{k}
        \left(\frac{2}{n\sigma}\right)^k	
        \left(\frac{2}{n'\sigma}\right)^k	
	\left(1-\frac{2}{n\sigma}\right)^{n-l-1-k}
	\left(1-\frac{2}{n'\sigma}\right)^{n'-l-1-k},
\end{multline}
Numerical computation of matrix elements via~\ref{elementtoy} is much faster than using the integral form of~\ref{elementintegraltoy}.
We also note that we can study the problem independently for each pair of $l,m$, because states corresponding to different values of these quantum numbers do not mix, as we can see from~\eqref{matrixtoy}, where Kronecker delta terms in indexes $l,l'$ and $m',m'$ are present.
For numerical diagonalization, we truncate the Hamiltonian matrix~\ref{matrixtoy} by restricting the number of basis vectors~\eqref{wfctstoy} - we take $n\in\{l+1,l+2,\dots,N\}$ for some suitable value of $N$.
Diagonalizing the truncated matrix gives the approximate ground state energy eigenvalues which we plotted in figure~\ref{fig.toy}.
%
\subsection{Exponential basis}\label{sec.toy.exp}
%
If we instead use the (non-orthogonal) basis set of exponential functions
\begin{equation}\label{exp-basis-toy}
	\langle \vec{x}|n \rangle = 2\alpha_n^{3/2} e^{-\alpha_n r} ,  
\quad n\in\{1,2,\dots,N\},
\end{equation}
where $\{\alpha_n\}_{n=1}^N$ is some suitable set of parameters,
we get (for $l=0$) the matrix elements of the Hamiltonian~\eqref{hamtoy}:
%
\begin{equation}\label{matrix-basis2-toy}
	\langle m |\hat{H}|n \rangle = 
       -(\alpha_m+\alpha_n-\alpha_m\alpha_n) \langle m|n \rangle 
       -g\frac{(2\sqrt{\alpha_m\alpha_n})^3}{(\alpha_m+\alpha_n+\mu)^2},
\end{equation}
where $\langle m|n \rangle$ are the overlap integrals obtained from~\eqref{exp-basis-toy} as
\begin{equation}\label{overlap-basis2}
	\langle m|n \rangle
	= 4(\alpha_m\alpha_n)^{3/2}\int_0^\infty \dd r r^2 e^{-(\alpha_m+\alpha_n)r}
	=\left(\frac{2\sqrt{\alpha_m\alpha_n}}{\alpha_m+\alpha_n}\right)^3.
\end{equation}
We choose the parameter set 
\begin{equation}\label{alphas}
	\alpha_n=A B^{n-1}, \qquad n\in {1,\dots, N},
 \end{equation}
where we used values $A=0.1$, $B=1.5$, after optimizing "by hand". 
The ground state energy, obtained as a results of Hamiltonian diagonalization, is shown again in Fig.~\ref{fig.toy}.

\subsection{Wronskian method}\label{sec.wronskian}
In this simple case of a Coulomb plus a Yukawa-type interaction in one dimension, it is easy to verify the calculations by supplementing them by a method based on the use of Wronskians. Let us briefly describe it.
The usual substitution $R(r)=u(r)/r$ in the Schr\"odinger equation corresponding to Hamiltonian~\eqref{hamtoy} gives:
\begin{equation}\label{rad.eqn}
	u''(r)=\left(k^2+\frac{l(l+1)}{r^2}-\frac2{r}-\frac{2e^{-\mu r}}{r}\right) u(r),
\end{equation}
with the energy eigenvalue $E=-k^2$. For small $r$, the solution is proportional to $r^{l+1}$, the asymptotic behavior at large $r$ is $e^{-kr}$, \cf~\cite[p.118]{landau}.
This motivates us to define two functions $u_{in}(r), u_{out}(r)$, being the solutions of~\ref{rad.eqn} and  satisfying the initial conditions (considering only $l=0$): 
\begin{equation}\label{init}
u_{out}(r_1)=r_1,\; u_{in}'(r_1)=1, \qquad
u_{out}(r_3)=e^{-k r_3},\; u_{out}'(r_3)=-k u_{out}(r_3). 
\end{equation}
\begin{wrapfigure}{R}{0.45\textwidth}
 \centering
   \vskip -3mm
                \includegraphics[width=0.45\textwidth]{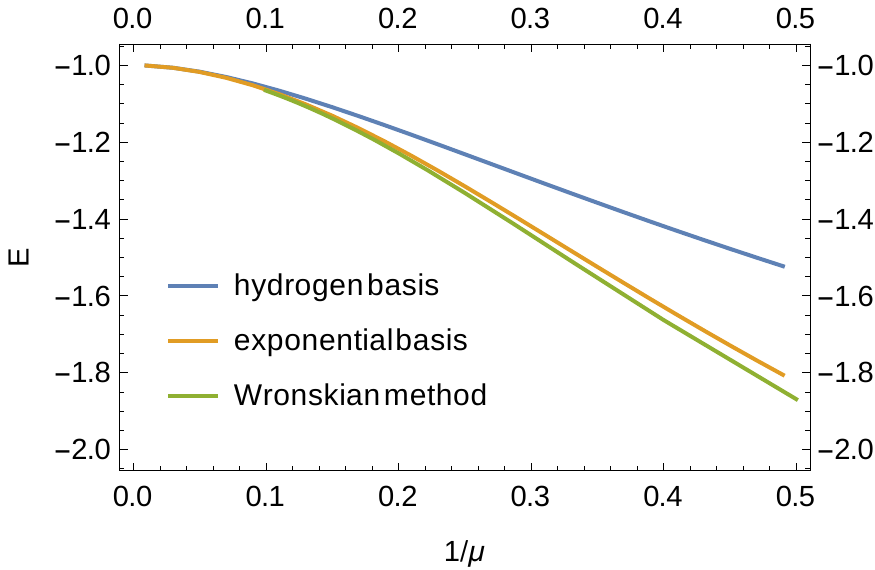}
   \vskip -1mm
   \caption{Toy model: ground state energy as a function of $1/\mu$.}
   \vskip -2mm
   \label{fig.toy}
\end{wrapfigure}
Let us note that the error of the method will be proportional to the error we make by imposing the initial conditions~\eqref{init} at finite distances (and we can estimate the error by varying the initial conditions).
 
To find the energy $E=-k^2$ (for a given $\mu$) of the ground state, we continue as follows: 
we define the Wronskian $W(r_2)=u_{out}(r_2)u_{in}'(r_2)-u_{out}'(r_2)u_{in}(r_2)$, where $u_{in}(r_2)$, $u_{out}(r_2)$ are function values at some intermediate point $r_2$ ($r_1<r_2<r_3$), obtained by numerical integration of the respective initial value problems.
We "sew" the two solutions $u_{in}(r)$, $u_{out}(r)$ by requiring them to be linearly dependent through the condition $W(u_{in}(r_2),u_{out}(r_2))=0$.  
The requirement of vanishing of the Wronskian at only one point $r_2$ is sufficient, because it implies its vanishing for all $r$.
This can been seen from the fact that $W'(u_{in},u_{out})=u_{out}u''_{in}-u_{out}''u_{in}=0$, as can be readily checked by inserting $u''_{in}$ and $u''_{out}$ from~\eqref{rad.eqn}.

We can see from figure~\ref{fig.toy} that for larger values of~$1/\mu$, the expansion on hydrogen atom basis does not follow the Wronskian method calculation as good as the expansion on the exponential functions.
%
This may be an indication that the basis set of exponential functions will be a better choice then the hydrogen atom basis also for the studied potential of~\eqref{ham}, consisting of a sum of Yukawa terms.
%
%
\section{Energy levels of the hydrogen atom in~$\R^3\times S^1$}
%
We shall now use the two basis sets of Secs~\ref{sec.toy.H} and~\ref{sec.toy.exp} to construct suitable bases for diagonalizing the Hamiltonian~\ref{ham}, defined on space~$\R^3\times S^1$. The periodicity of space along the compactified dimensions means that the wave function is periodic in the fourth coordinate $\theta$. This suggests to take plane waves as a basis of~$S^1$.
%
\subsection{Basis constructed from the hydrogen atom eigenstates}
%
We now use the basis set~\ref{wfctstoy} to construct a basis of space~$\R^3\times S^1$ 
 $\R^3$ by adding plane waves along the extra dimension:
\begin{equation}\label{wfcts}
	\langle \vec{x}|nlmq \rangle = R_{nl}(r)Y_{lm}(\Omega)\frac{e^{\ii q \theta}}{\sqrt{2\pi}}, \quad l\in\N,\, m\in\{-l,\dots,l\},\, n\in\{l+1,l+2,\dots\},\, q\in \Z.
\end{equation}
Here $q$ is a new quantum number that specifies the particular Fourier mode along the compactified dimension.
Calculation of the matrix elements from \eqref{ham} and~\eqref{wfcts} gives:
\begin{equation}\label{matrix}
	\langle n'l'm'q' |\hat{H}|nlmq \rangle = \delta_{ll'} \delta_{mm'}
	\left \{ \delta_{nn'} \delta_{qq'}\left(-\frac1{n^2}+\frac{q^2}{R^2}\right)- 
	\left(1-\delta_{qq'}\right) M_{n,n';l}(1,|q-q'|/R) \right \},
\end{equation}
where $M_{n,n';l}(1,|q-q'|/R)$ is given by~\eqref{elementtoy} and consequently $\sigma(|q-q'|/R)=1/n+1/n'+|q-q'|/R$.
Here $\left(1-\delta_{qq'}\right)$ indicates that the corresponding term is present only for $q\neq q'$.
As in Sec.~\ref{sec.toy.H}, we can study the problem independently for each pair of quantum numbers $l,m$. 
For numerical diagonalization, we truncate the Hamiltonian matrix by restricting quantum numbers $n$ and $q$ to ranges:
\begin{equation}
\quad n\in\{l+1,l+2,\dots,N\},\; q\in\{-Q,\dots,0,\dots Q\},
\end{equation}
where $N$ and $Q$ define the number of basis vectors in the truncated basis.
We note that, as known from the properties of Fourier series, in the limit $Q\to\infty$ the plane-wave basis becomes a complete basis set (here for functions on $S^1$). This is not true for the basis of hydrogen atom bound states which, in the limit $N\to\infty$, remains an incomplete basis set for functions on space $\R^3$ (the eigenstates corresponding to the continuous spectrum are missing).
 
Diagonalizing the truncated Hamiltonian matrix~\eqref{matrix} numerically using Mathematica, we obtained the approximate energy eigenvalues which we plotted in figure~\ref{fig:energy}.
The lifting of degeneracy of energy levels $n=2,3,4$ for $l=0$ and $l=1$ is shown in figure~\ref{fig:lifting}. 

The approximate solution of the problem, \ie~the eigenvector obtained as a result of Hamiltonian diagonalization in the (truncated) basis constructed from hydrogen atom eigenstates, is 
\begin{equation}\label{H.fct}
	\psi_{lm}(\vec{x})=
	\sum_{n=1}^N\sum_{q=-Q}^Q
	a_{n,q} R_{nl}(r)Y_{lm}(\Omega)\frac{e^{\ii q \theta}}{\sqrt{2\pi}}, 
\end{equation}
where $a_{n,q}$ are the components of the (normalized) eigenvector (weights of corresponding basis states) and $N,Q$ define the size of the truncated basis.
The probability density $|\psi_{lm}{(r,\theta)}|^2$ of finding the electron at point $(r,\theta)$ is plotted in figure~\ref{fig:probab} for several compactification radii.
The results will be discussed in the concluding Sec.~\ref{conclusion}.
\subsection{Basis constructed from exponential functions}
Let us now modify the basis~\eqref{exp-basis-toy}. 
Considering only spherically symmetric states ($l=0$),%
%
%
let us repeat the calculations by using a (non-orthogonal) set of basis states constructed from exponential functions: 
\begin{equation}\label{exp-basis}
	\langle \vec{x}|iq \rangle = 2\alpha_i^{3/2} e^{-\alpha_i r} \frac{e^{\ii q \theta}}{\sqrt{2\pi}},  
\quad i\in\{1,2,\dots,I\},\; q\in\{-Q,\dots,0,\dots Q\}
\end{equation}
where the parameter set $\{\alpha_i\}_i$ is kept as in~\eqref{alphas}.
The matrix elements of the Hamiltonian~\ref{ham} in this basis are:
\begin{equation}\label{matrix-basis2}
	\langle jp |\hat{H}|iq \rangle = 
      \left[ \langle jp|iq \rangle 
	      \left(\alpha_i\alpha_j+\frac{q^2}{R^2} \right)
	-\frac{(2\sqrt{\alpha_i\alpha_j})^3}{(\alpha_i+\alpha_j+|q-p|/R)^2}
      \right].
\end{equation}
where
\begin{equation}\label{overlap-basis2}
	\langle jp|iq \rangle = 
	4(\alpha_m\alpha_n)^{3/2}\int_0^\infty \dd r r^2 e^{-(\alpha_i+\alpha_j)r} \frac1{2\pi}\int_0^{2\pi}e^{\ii (q-p) \theta}
	=\left(\frac{2\sqrt{\alpha_i\alpha_j}}{\alpha_i+\alpha_j}\right)^3 \delta_{p,q}.
\end{equation}
are the overlap integrals obtained from~\eqref{exp-basis}.
The results of Hamiltonian diagonalization are plotted in figure~\ref{fig:energy}, Fig.~\ref{fig:probab} then shows the probability density, obtained as $|\psi_{lm}{(r,\theta)}|^2$ from an analogical expression to~\eqref{H.fct}.
%
\section{Comparison of methods, interpretation, summary}\label{conclusion}
\begin{figure}[!htb]
	\centering
	\begin{subfigure}[b]{0.47\textwidth}
               \includegraphics[width=\textwidth]{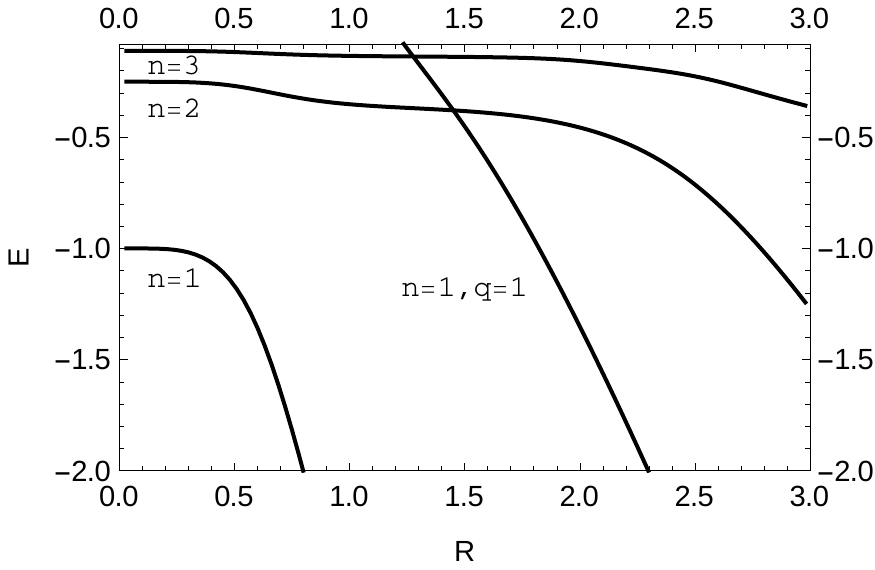}
	       \caption{Energy levels: hydrogen atom basis\\ 
		       ($l=m=0$, size $N=10$, $Q=30$)}
        \end{subfigure}%
	\hspace{0.15cm}
	\begin{subfigure}[b]{0.47\textwidth}
               \includegraphics[width=\textwidth]{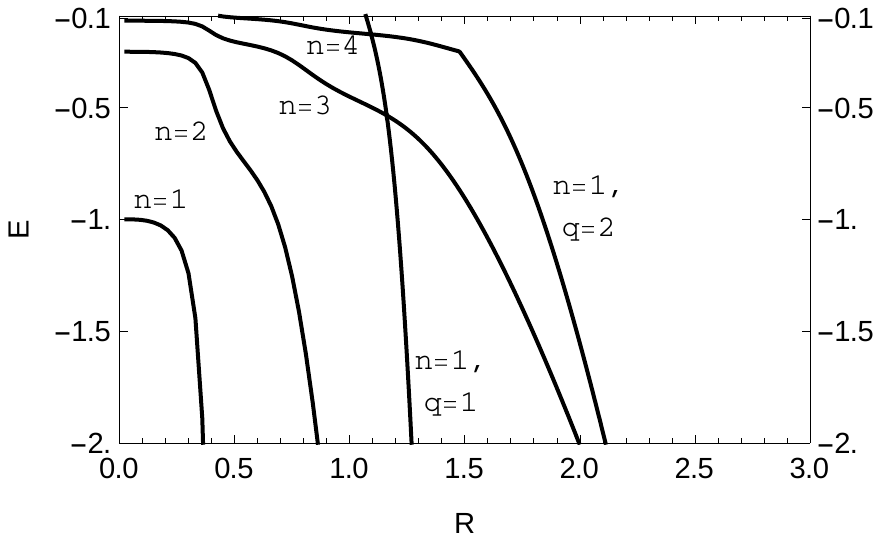}
	       \caption{Energy levels: exponential basis\\ 
		       ($l=m=0$, size $N=10$, $Q=30$)}
        \end{subfigure}%
\qquad
\vskip0.5cm
	\begin{subfigure}[b]{0.47\textwidth}
               \includegraphics[width=\textwidth]{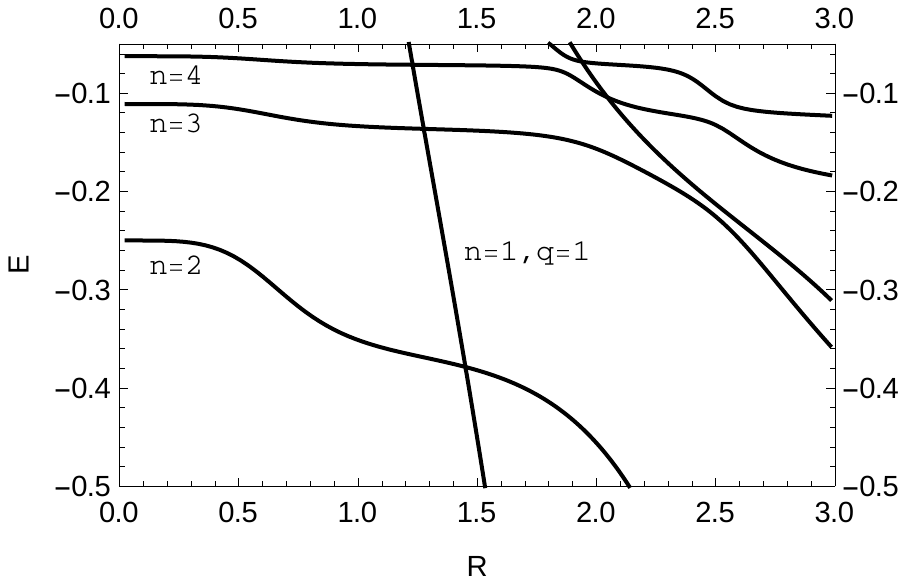}
	       \caption{Energy levels: hydrogen atom basis\\ 
		       ($l=m=0$, size $N=10$, $Q=30$)}
        \end{subfigure}%
	\hspace{0.15cm}
	\begin{subfigure}[b]{0.47\textwidth}
               \includegraphics[width=\textwidth]{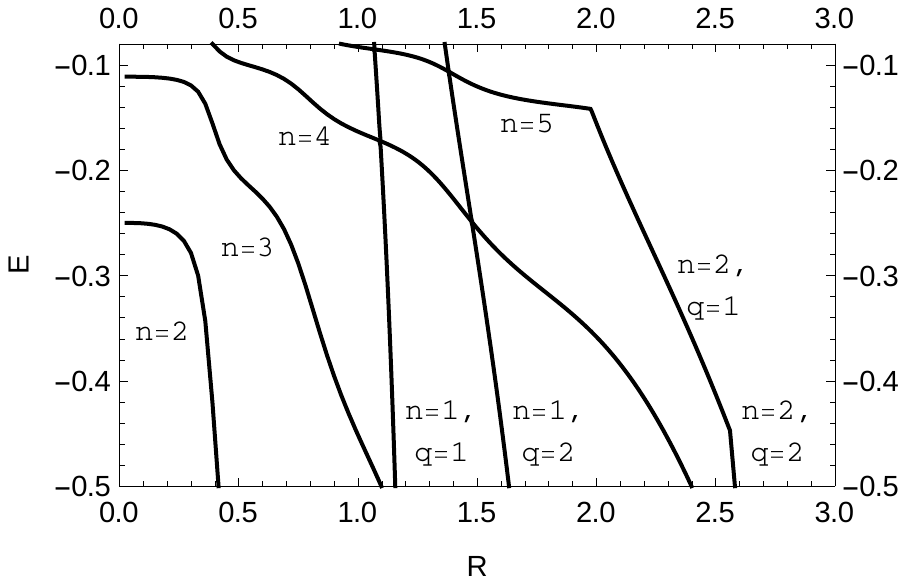}
	       \caption{Energy levels: exponential basis\\ 
		       ($l=m=0$, size $N=10$, $Q=30$)}
        \end{subfigure}%
\qquad
\vskip0.5cm
	\begin{subfigure}[b]{0.44\textwidth}
             \includegraphics[width=\textwidth]{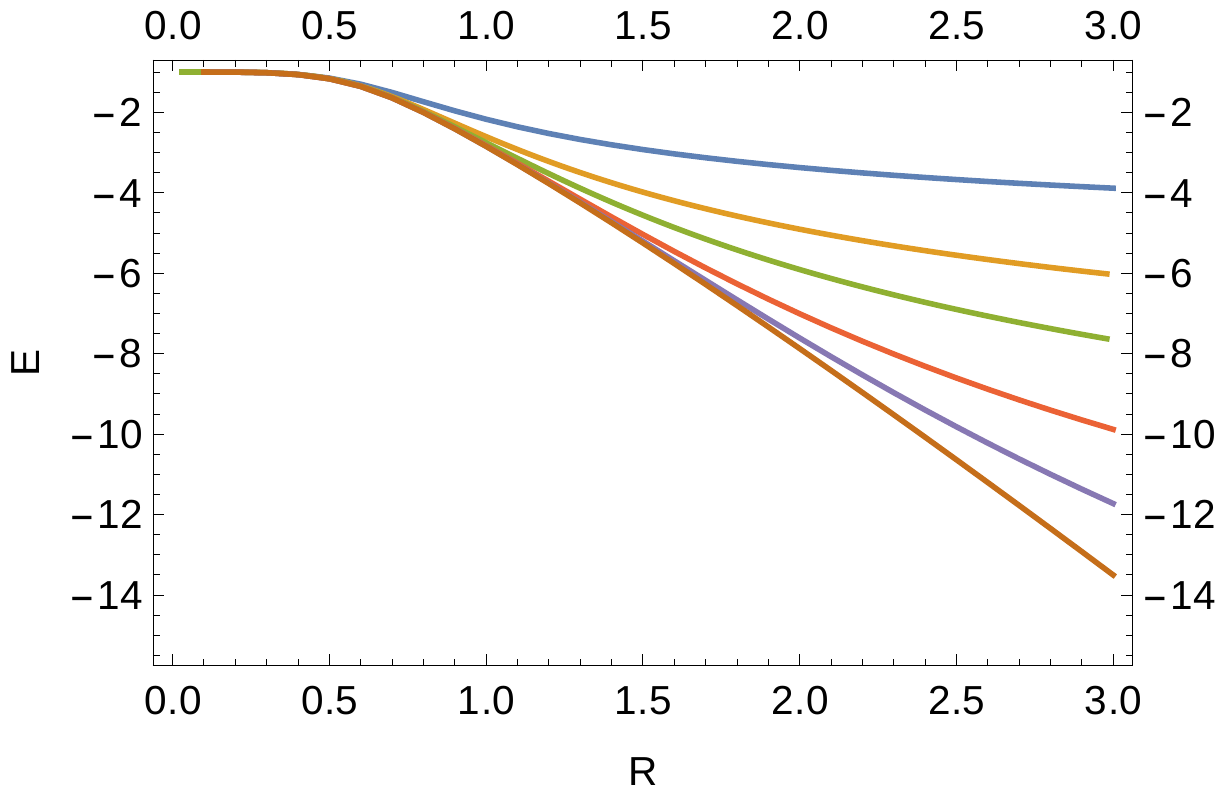}
	     \caption{Ground state energy dependence on basis size\\
	     (hydrogen atom basis, $N=7$, $Q=1,\dots,50$)}
        \end{subfigure}%
	\hspace{0.22cm}
	\begin{subfigure}[b]{0.08\textwidth}
               \includegraphics[width=\textwidth]{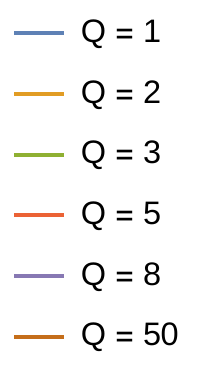}
	       \vspace{1.7cm}
        \end{subfigure}%
	\hspace{0.0cm}
	\begin{subfigure}[b]{0.44\textwidth}
               \includegraphics[width=\textwidth]{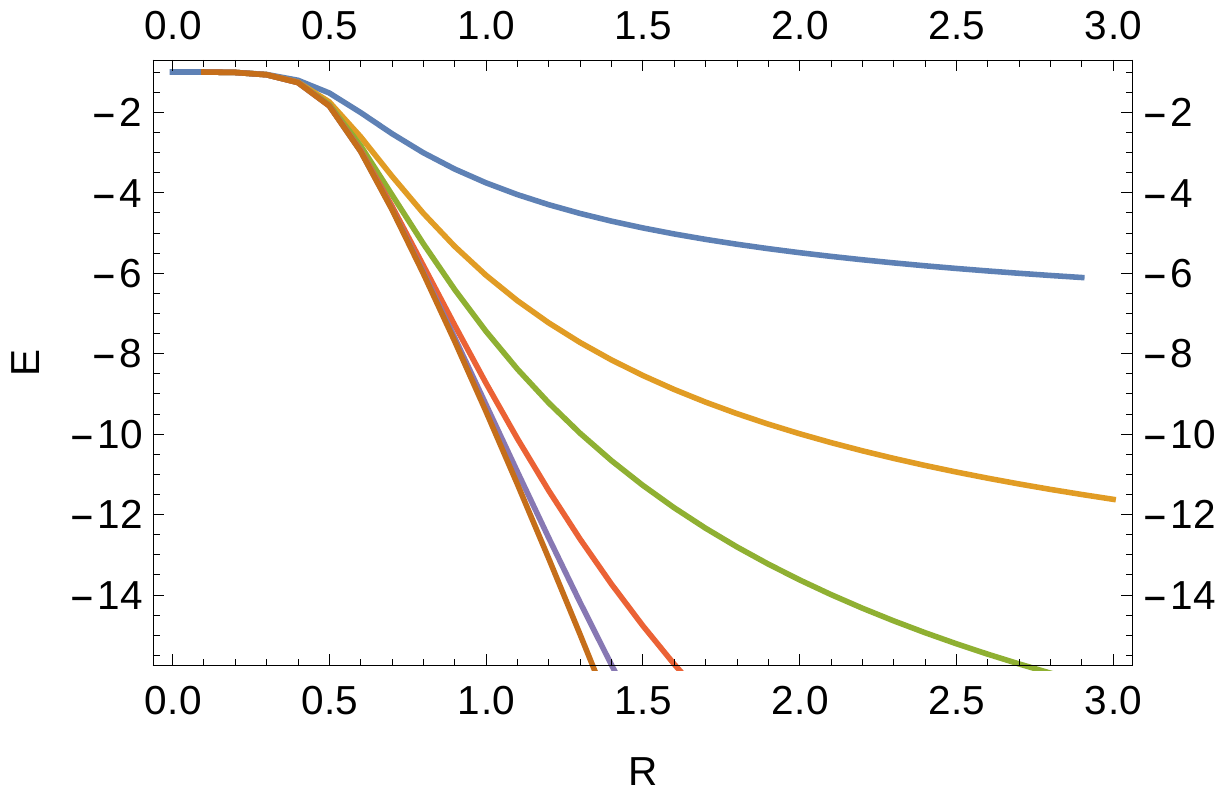}
	     \caption{Ground state energy dependence on basis size\\
	     (exponential basis, $N=7$, $Q=1,\dots,50$)}
        \end{subfigure}%
\caption{Energy eigenvalues (in units $e^2/2a$) as a function of the compactification radius $R$ (in units of the Bohr radius $a$).
Hydrogen atom basis (left-hand side), exponential basis (right-hand side).
The computational step in $R$ was adjusted according to the second derivative of the curves between $\Delta R=0.005$ and $\Delta R=0.03$.}
                \label{fig:energy}
\end{figure}

%
Let us now analyze the results. In Fig.~\ref{fig:energy}, we plotted the calculated spectrum, as a function of the compactification radius, for the lowest energy levels.
The plots on the left-hand side were obtained in basis~\eqref{wfcts} constructed from the bound states functions of the hydrogen atom,
whereas the right-hand side refers to basis~\eqref{exp-basis}, constructed from exponential functions.
Comparing Figures (A) to (B), or (C) to (D), we can see that the exponential basis provides lower energies than the hydrogen atom basis.
This confirms our expectations from the toy model of a single Yukawa potential, \cf~Sec.~\ref{sec.toy}. 
There, the calculations in the hydrogen atom basis could not follow the curve obtained from the Wronskian method as good as the exponential basis.

Energy levels with quantum numbers $q=0$ correspond to electron states which do not possess any $\theta$-dependence of the wave function (their wave functions are constant in the compactified direction).
Levels with $q\geq 1$ are the excited Kaluza-Klein states. 
As the compactification radius $R$ increases, their energy decreases as $1/R^2$, \cf~\eqref{matrix} or~\eqref{matrix-basis2}. 
For $R$ large enough, their energy falls down to a range comparable to the bound spectrum and they cross levels with $q=0$, \cf~plots (A-D) in Fig.~\ref{fig:energy}.
We note that on the right-hand side of plots (B) and (D), some of the levels seem to change direction abruptly. This is a mere consequence of the fact that for each $R$ we only plot a certain (fixed) number of lowest eigenvalues, hence the continuations after the "break" are not the same levels but the excited Kaluza-Klein states (with $q\geq 1$, see above).

It was shown in~\cite{Bures-Siegl-2015} that the hydrogen atom with an additional compactified dimension becomes unstable for $R_{crit}>a/4$. 
Calculations using basis~\eqref{wfcts} give for $R=a/4$ the ground state energy $E_1=-1.00665$,
the exponential basis gives $E_1=-1.08315$ (both results for $N=10$, $Q=30$).
Figs.~\ref{fig:energy}, parts (E) and (F), show the dependence of the ground state energy on the respective bases sizes.
When truncating to a finite basis, we do not get a clear sign of divergence for compactification radii close to the critical value $R_{crit}$. 
However, we have confirmed the that the spectrum changes dramatically, when we approach the critical compactification radius $R_{crit}$.
For small compactification radii, we can trust our results which agree between the different methods and do not depend on the size of the basis.
For a more precise calculation of the spectrum close to $R_{crit}$ one would have to use a different technique.

\begin{figure}[!htb]
	\centering
       \begin{subfigure}[b]{0.5\textwidth}
            \includegraphics[width=\textwidth]{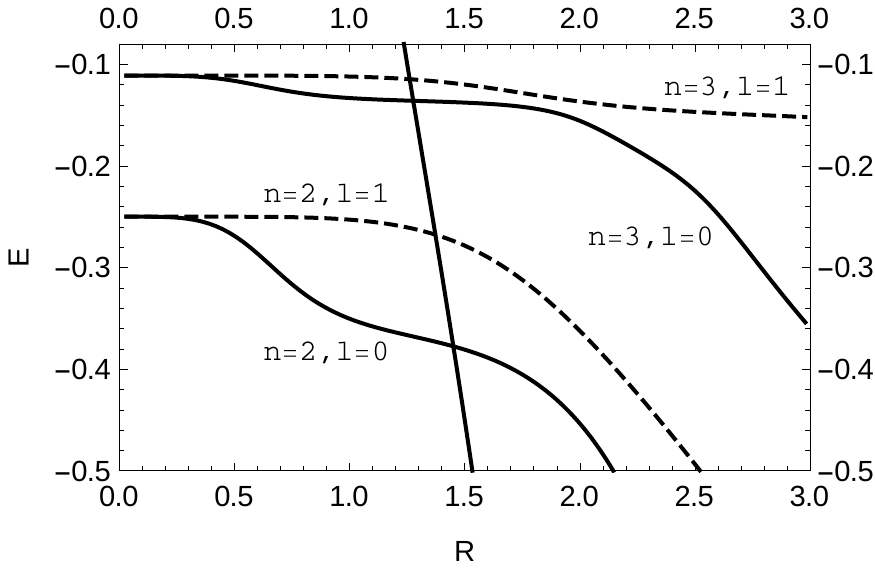}
       \end{subfigure}%
       \caption{Lifting of degeneracy of energy levels (hydrogen atom basis, $N=7$, $Q=30$): 
	        $n=\{2,3\}$: $l=0$ (solid line) $l=1$ (dashed line), $m=0$.
	        The almost vertical curve represents the first Kaluza-Klein state $n=1,q=1$.}
       \label{fig:lifting}
\end{figure}

In Fig.~\ref{fig:lifting}, we can see how the presence of an extra dimensions lifts the degeneracy of levels $n=2,3$ for quantum numbers $l=0$ and $l=1$ (the $n=1$, $q=1$ Kaluza-Klein state is also visible). This is an effect analogous to the splitting of the spectral lines of atoms due to various corrections (\eg~fine or hyperfine structure).
Recent progress in spectroscopic methods could serve as a testing ground providing further constraints on the size of extra dimensions~\cite{1367-2630-17-3-033015}, as the effects predicted by theory can be directly measured:
comparing the calculated spectrum to experimental values would restrict the possible values of the compactification radius $R$.

\begin{figure}[!htb]
	\centering
	\begin{subfigure}[b]{\textwidth}
               \includegraphics[width=\textwidth]{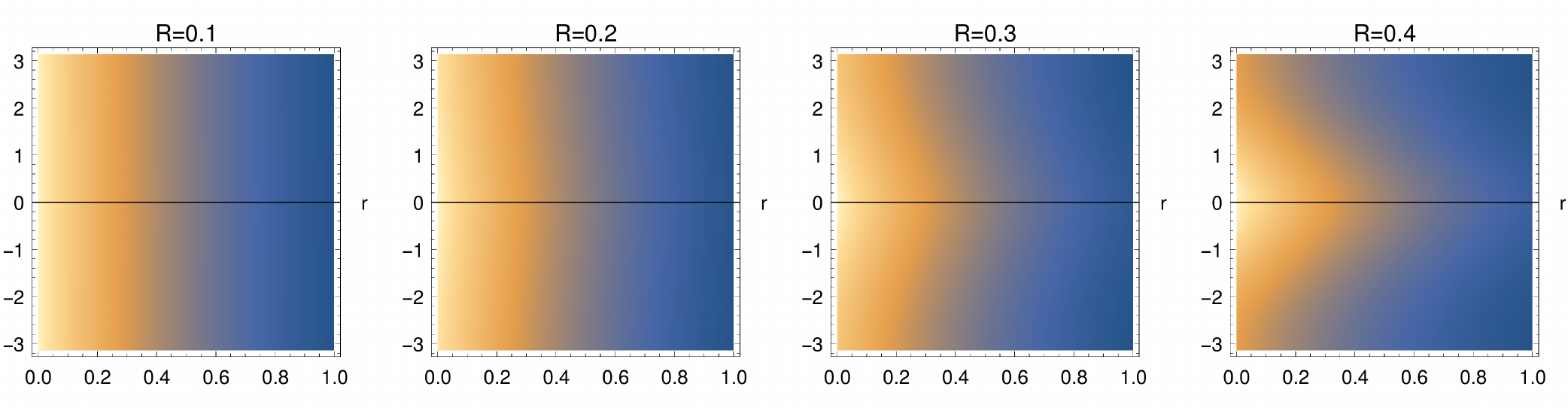}
	\caption{Hydrogen atom basis ($N=10$, $Q=30$)}
        \end{subfigure}%
	\\
	\begin{subfigure}[b]{\textwidth}
               \includegraphics[width=\textwidth]{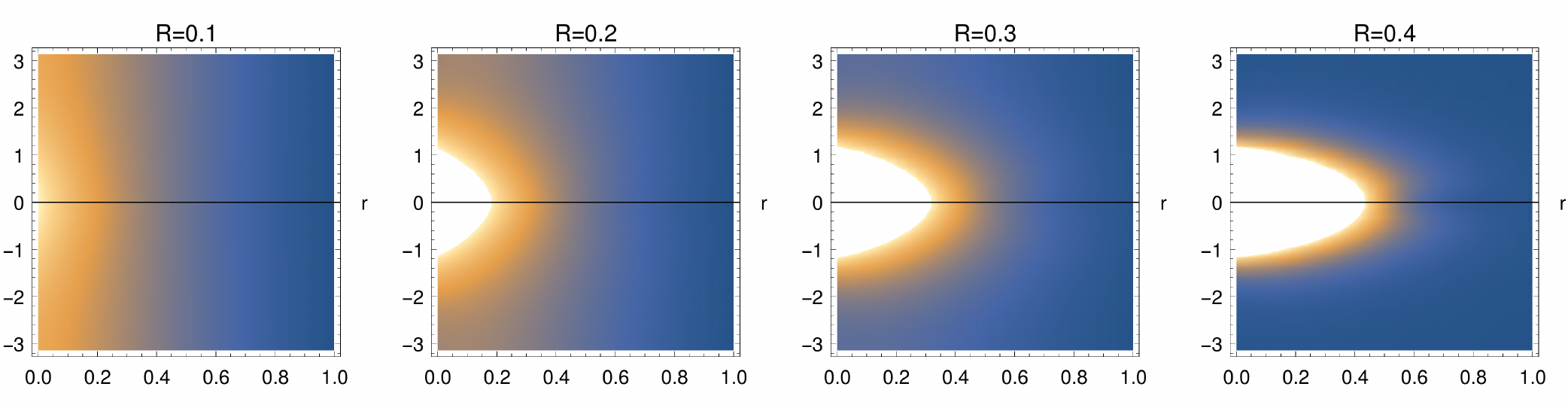}
	       \caption{Exponential basis ($N=10$, $Q=30$)}
        \end{subfigure}%
	\caption{Probability density in the $(r,\theta)$ plane for several values of $R$.}
	\label{fig:probab}
\end{figure}

Finally, Fig.~\ref{fig:probab} shows the electron probability density, as calculated by using the two sets of basis vectors.
The top row corresponds to basis~\eqref{wfcts} constructed from bound states functions of the hydrogen atom,
whereas the bottom row is based on calculations in basis~\eqref{exp-basis}, constructed from exponential functions.
We can observe how the electron tends to localize at the nucleus, as the compactification radius increases.
To achieve localization, Fourier modes along the compactified direction are required. 
However these terms increase the energy of the electron (which is proportional to $1/R^2$, \cf~\eqref{matrix} or~\eqref{matrix-basis2}), 
so for a given value of $R$, some sort of balance has to be found.
Again, for large enough compactification radius, close to $R_{crit}=a/4$, \cf~\cite{Bures-Siegl-2015}, complete localization or the eventual "fall to the center", \cf~\cite[p.53-54,114-177]{landau}, is expected. 

The present calculational method (partial exact diagonalization) provides an upper bound on the real energy. Hence, it would be of great interest to use some other method to obtain also a lower bound on the spectrum, and to better explore the behavior of the spectrum close to the critical compactification radius.

{\footnotesize \bibliographystyle{ieeetr}}

\end{document}